\documentclass[prb,twocolumn,showpacs,preprintnumbers,amsmath,amssymb]{revtex4}
\usepackage{dcolumn}
\usepackage{bm}
\usepackage{graphicx}
\usepackage{bm}
\begin{document}
\pacs{75.47.Lx, 75.50.Lk, 75.30.Kz}
\title{Finite size effects and the effect of disorder on electronic phase separation in    Pr${_{0.5}}$Ca${_{0.5}}$Mn${_{1-x}}$Al${_{x}}$O${_3}$}
\author{Sunil Nair}
\altaffiliation{Present address: Department of Condensed Matter Physics and Materials Science; Tata Institute of Fundamental Research, Mumbai 400005, India}
\author{A. Banerjee}
\email{alok@csr.ernet.in}
\affiliation{UGC-DAE Consortium for Scientific Research,\cite{IUC}\\University Campus, Khandwa Road, Indore, 452 017, INDIA.}
\date{\today}
\begin{abstract}
We report a bulk magnetisation study of the low temperature magnetic phases in the Mn site substituted half doped manganite series  Pr${_{0.5}}$Ca${_{0.5}}$Mn${_{1-x}}$Al${_{x}}$O${_3}$. Our results indicate the formation of self organised structures in this series of compounds which is driven by electronic phase separation and stabilised by the presence of finite quenched disorder, thus providing a basis for understanding the anomalously large supression of the CE phase in the presence of disorder. Striking magnetic memory effects similar to that seen in spin glasses and interacting superparamagnets are also observed.  Increasing disorder is seen to make electronic phase separation weaker owing to the destabilisation of the CE state which results in an enhanced mobility of the charge carriers. 
\end{abstract}
\maketitle 
The propensity of certain transition metal oxides (TMO) like cuprates, manganites and nickelates to remain insulating in regions of their phase diagram inspite of finite hole doping continues to be a subject of intense scrutiny. These systems remain insulating due to the rather strong electron correlations which prevent the formation of a metallic ground state. In some systems, this is achieved by the formation of self organised inhomogenities which subdivide the system into regions with varying hole concentrations, a phenomenon which in common parlance is referred to as electronic phase separation (EPS).\cite{dag1} An interesting example of this phase separation in 3D systems is the formation of \emph{charged domain walls} \cite{zaanen} which occur as a compromise between the antiferromagnetic interaction between ions and the Coulombic interaction, both of which favour localisation of charges vis a vis the kinetic energy of the doped carrier, which tends to delocalise charges. Along this wall the charge density is constant except for the modulation imposed by the lattice, whereas perpendicular to it, the charge density varies drastically.\cite{tranquada} In many practical systems, though the electron overlap may not be large enough as compared to the Coulombic repulsion to prevent the formation of this structure, it can cause their positions to fluctuate rapidly as compared to the measurement time scales, thus making experimental investigations non trivial.\cite{kivelson}

Though the existance of charged domain walls have been unambigously established in cuprates\cite{tranquada, tranquada1} and nickelates,\cite{tranquada2} the observation of such a state in manganites which arise purely out of EPS remains nonconclusive. In manganites, a self organised structure of this nature would be expected to arise either in the lightly doped ($\leq 10\%$) or the over doped ($\geq 50\%$) regions of the phase diagaram, since in these regions the electronic ground state is observed to be insulating. Needless to say, it is in these concentrations that self organised structures like an orbital domain state \cite{domain}(which comprises of Ferromagnetic insulating domains separated by ferromagnetic metallic walls) and charge ordered, Jahn Teller stripes \cite{jt}(which arises due to minimisation of the strain energy associated with the Jahn Teller distorted Mn${^{3+}}$O${_6}$ octahedra) have been observed. Here, we report the observation of finite size effects in a series of Mn site substituted half doped manganite Pr$_{0.5}$Ca$_{0.5}$Mn$_{1-x}$Al$_{x}$O${_3}$ and argue that these drastic effects can be explained to arise due to the formation of self organised structures with varying hole concentrations. Not only does this bring forward the difference in both the nature of EPS and the dynamics of the EPS clusters as a function of substitutional disorder in the half doped manganites vis a vis other TMO systems, but also provides a natural basis for understanding the profound susceptibility of the CE phase to quenched disorder.

The parent Pr${_{0.5}}$Ca${_{0.5}}$MnO${_3}$ is a narrow bandwidth system, where a robust charge and orbital ordering is seen to set in around 240K.\cite{jirak} Here, the magnetic ordering is known to be of the so called CE type , which is made up of checkerboard arrangement of Mn${^{3+}}$ and Mn${^{4+}}$ ions, accompanied by an associated orbital ordering at the Mn${^{3+}}$ sites, where the d${_{3x{^2}-r{^2}}}$/d${_{3y{^2}-r{^2}}}$ orbitals order alternatively in the \emph{ab} plane.\cite{goodenough} Interestingly, the correlation lengths for the Orbital Order (OO) is seen to be different than that of the CO (with the OO being shorter than that of the CO), indicating the presence of an orbital domain state.\cite{zim} Finite amounts of disorder can be introduced in the host Mn lattice by impurity substitution leading to a fascinating array of properties induced as a consequence of both the bandwidth of the parent compound as well as the electronic configuration of the impurity used.\cite{dho, kimura, mahend} In the reported system, Mn is judiciously substituted by Al on ionic considerations as well as on account of its non magnetic nature to achieve random disorder in the magnetic sublattice without adding any lattice distortions or dopant mediated magnetic interactions. Polycrystalline samples prepared by the solid state ceramic route were used, and the details of the sample preparation and characterisation can be found elsewhere.\cite{sunil}

\begin{figure}[htb]
	\centering
		\includegraphics[width = 8 cm]{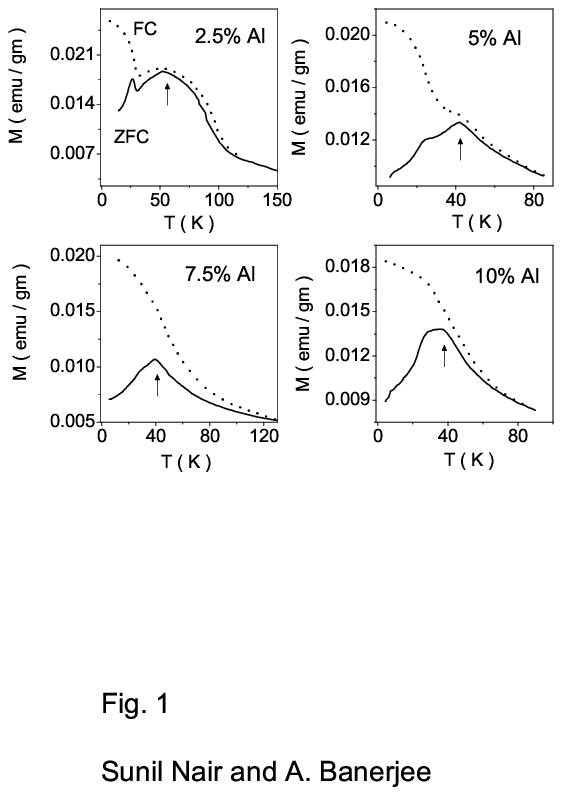}
	\caption{Bulk DC magnetisation measurements as a function of temperature for the series   Pr${_{0.5}}$Ca${_{0.5}}$Mn${_{1-x}}$Al${_{x}}$O${_3}$ (with x = 2.5, 5, 7.5 and 10\%)at a measuring field of 100 Oe. The continuous line is for zero-field cooled and the dotted line is for field cooled branches. The peak in magnetisation associated with the AFM transition (as marked by an arrow) is seen to reduce with increasing Al substitution.}
	\label{Fig1}
\end{figure}

\begin{figure}[htb]
	\centering
		\includegraphics[width = 8 cm]{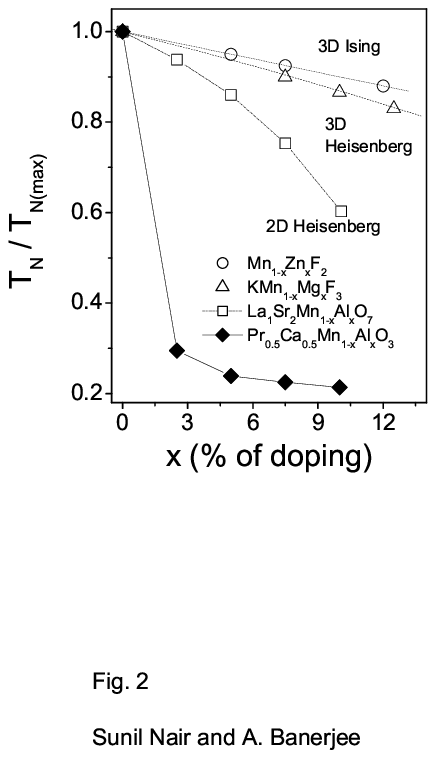}
	\caption{The normalised Neel temperatures (T${_N}$/T${_{max}}$) plotted as a function of non magnetic substitution for various 3D magnetic systems. The rapid fall of T${_N}$ in the system Pr${_{0.5}}$Ca${_{0.5}}$Mn${_{1-x}}$Al${_{x}}$O${_3}$ is clearly seen. The data for the other samples have been taken from Reference 19. The skewed lines are guides to the eye indicating the effect of site dilution of the different magnetic systems.}
	\label{Fig2}
\end{figure}

 An interesting manifestation of the disorder introduced by the substitution of Al is the observation of a low temperature metastable ground state, which has been shown to arise from the superparamagnetic like \emph{blocking} of antiferromagnetic (AFM) clusters.\cite{sunil1} However, the most drastic effect of Al susbtitution is the anamolously large supression of the AFM transition temperature (T${_{N}}$). The parent Pr${_{0.5}}$Ca${_{0.5}}$MnO${_3}$ is known to have a long range AFM T${_N}$ $\approx$ 170K, which drops to $\approx$ 50K on the introduction of 2.5$\%$ of Al substitution. Introduction of relatively small amounts of Al substitution (upto 10$\%$)is seen to result in a substantial modification of the low temperature magnetic properties, as is shown in Figure ~\ref{Fig1}. 
Though this rapid suppression of the T${_N}$ of CE systems in the presence of disorder has been predicted using Monte Carlo Simulations\cite{dag2} and has been experimentally observed in narrow bandwith half doped systems in the presence of Mn site substitution,\cite{damay} adequate attention has not been paid to either the underlying physics or the true determination of the magnetic ground state. As is shown in Figure ~\ref{Fig2}, the rate of fall of T${_N}$ as a function of Al substitution in Pr$_{0.5}$Ca$_{0.5}$Mn$_{1-x}$Al$_{x}$O${_3}$ is much larger than that observed in site diluted Heisenberg or Ising systems \cite{book} and thus cannot be explained on the basis of simple site dilution alone.
It is important to note here that the broader bandwidth bilayered La${_1}$Sr${_2}$Mn${_{2-x}}$Al${_x}$O${_3}$ system shows a systematic reduction in T${_N}$ as is expected for site diluted 2D Hesisenberg systems. \cite{sunil3}
 
Since the suppression of T${_{N}}$ cannot be accounted for by a simple dilution of the magnetic lattice, we argue that this observation is a consequence of finite size effects in this Mn site substituted system. Though the subdivision of a long range magnetically ordered system into domains is reasonably well understood in the case of ferromagnets; its energetics as far as antiferromagnets are concerned is more complex. While a long range ferromagnet breaks up into subdomains to minimise the magnetostatic energy arising due to the presence of demagnetising fields,\cite{aharony} in systems with AFM order the formation of a domain structure is closely related to the presence of anisotropies or crystal defects like dislocations, grain boundaries, impurities etc.\cite{jetp}  The classic work of Imry and Ma \cite{ma}(and more recently Burgy et al\cite{burgy}) have indicated that systems with quenched random fields (like dilute antiferromagnets, where a random field originates from the replacement of magnetic entities in an AFM compensated structure), the long range ordered state is unstable . In other words, the system would break up into domains of size \emph{L}, where \emph{L} is determined by the interplay between the domain wall energy and the statistics of the random fields. As is well known, the finite size scaling theory predicts that the measured T${_N}$ is limited by the size (\emph{L}) of the system in the form 1-T${_N}$(\emph{L})/T${_N}$($\infty$) $\propto$ \emph{L}$^{-1/\nu}$ (where $\nu$ is the correlation exponent); thus leading to a shift and rounding of the critical region in systems with finite (though microscopically large) dimensions.\cite{fss} Similar finte size effect is confirmed for Pr${_{0.5}}$Ca${_{0.5}}$Mn${_{0.975}}$Al${_{0.025}}$O${_3}$ and the size of the clusters are estimated from the analysis of linear and nonlinear susceptibility. \cite{sunil1} 
  
\begin{figure}[htb]
	\centering
		\includegraphics[width = 8 cm]{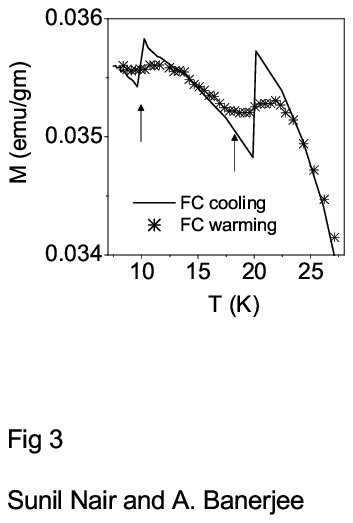}
	\caption{The magnetic memory effect observed in the Pr${_{0.5}}$Ca${_{0.5}}$Mn${_{0.975}}$Al${_{0.025}}$O${_3}$ sample. The measurements were done at an applied field of 1 kOe with stops of 20 minutes each at 10K and 20K respectively.}
	\label{Fig3}
\end{figure}

 An interesting manifestation of EPS in this finite size system is the pronounced hysteresis in magnetic susceptibility as a function of thermal cycling which occurs when the holes migrate from the bulk of the clusters to the domain walls without changing the size of these clusters.\cite{sunil1} It is to be noted that such hysteretic behaviour accompanied by a metastable ground state is now being reported for many other manganite systems in the presence of quenched disorder,\cite{fisher} indicating that this novel EPS is prevailent in a variety of narrow bandwidth manganite systems. The magnetic ground state of this system being metastable is associated with large relaxation times and gives rise to striking memory effects. This is shown in  Figure~\ref{Fig3}, where the Field cooled cooling and warming curve with a 
specific temperature and field protocol is shown.
In the Field cooled measurement cycle, the applied field was switched off and the temperature was held stable at 10 K and 20 K for a waiting time of 20 minutes each. On completion of the cooling cycle, a warming cycle is initiated at the same measurement field. As is clearly seen, in the warming cycle, distinct step like features are seen at the temperatures where the measurement was stopped in the cooling runs; thus showing that the system retains memory of its thermal history. Similar effects have been observed in some other manganite systems as well,\cite{levy} and is known to be typical of spin glasses\cite{sg} or interacting magnetic nano particle systems.\cite{sal} In the present system, these memory effects could arise as a result of inter cluster interactions or/and cluster size distributions.\cite{sal1,sal2}   
 
 A finite size scaling analysis of the susceptibility, where the susceptibility in the AFM regime can be scaled as $\chi(x, T) = \chi\left\{f(x)\left[T-T{_N}(x)\right]\right\}$, where x indicates the etent of hole doping has been used to indicate the formation of self organised structures in the lightly doped cuprates.\cite{cho, suh} However, a similar scaling procedure cannot be used in the series studied in this work, because the nature of structures formed as a consequence of EPS is quite different in the half doped manganites. This is primarily because of the fact that in the cuprates, the doped holes segregate into domain walls in such a fashion that the width of the walls is invariant with hole doping.\cite{cho} This would mean that the region enclosed within these walls reduce with increasing doping, thus resulting in substantial finite size effects as evidenced by a pronounced reduction in T${_N}$. In the system Pr${_{0.5}}$Ca${_{0.5}}$Mn${_{0.975}}$Al${_{0.025}}$O${_3}$, we have shown that thermal cycling results in a novel phase separation, where holes migrate to the domain walls from the bulk keeping the size of the clusters invariant, thus resulting in a lower T${_N}$ as dictated by the extent of hole migration and the phase diagram of the parent compound. Thus, whereas in the cuprates phase separation occurs between hole rich and hole free regions, in the Mn site substituted manganites EPS is much more subtle and results in regions with reasonably similar (CE and Pseudo CE) ground states.\cite{sunil1, yaicle}  
  
\begin{figure}[htb]
	\centering
		\includegraphics[width = 8 cm]{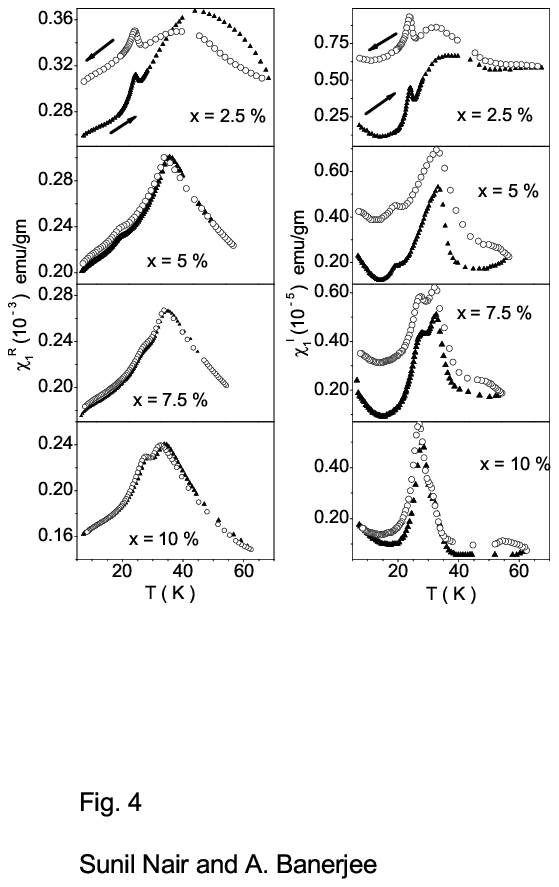}
	\caption{The temperature dependence of the real and imaginary parts of ac susceptibility as measured in the series Pr${_{0.5}}$Ca${_{0.5}}$Mn${_{1-x}}$Al${_{x}}$O${_3}$ for warming (filled traingles) and cooling (open circles) cycles. All the measurements were done at an applied field of 2.5 Oe and an exciting frequency of 733 Hz. The extent of thermal hysteresis is clearly seen to decrease with increasing Al substitution.}
	\label{Fig4}
\end{figure}

 In systems like the lightly doped cuprates, the electronic phase separation is dynamic in the sense that the antiphase domain walls (and hence the enclosed domains) are mobile. The presence of a small but finite disorder makes these self organised structures more static, thus enabling its observation in more convenient experimental time scales.\cite{kivelson} The effect of increasing disorder in these systems have been investigated,\cite{nh, hammel} and it is now understood that increasing disorder can either result in the pinning of the domain wall or in its evaporation due to the pinning of the constituent holes to charged impurities. Figure~\ref{Fig4} shows the hysteresis in the real and imaginary parts of the ac susceptibility in the series Pr$_{0.5}$Ca$_{0.5}$Mn$_{1-x}$Al$_{x}$O${_3}$.
As is seen, the extent of hysteresis (which in turn is an indication of the extent of EPS) consistently reduces as a function of increasing Al substitution. However, as was seen from Figure~\ref{Fig1}, there is no apparent relation between the amount of disorder and the blocking temperature (which remains almost independent of x). This would imply that there is no significant change in the size of the AFM clusters with increasing Al substitution.   
 
 This interesting observation of decreasing EPS with increasing disorder can be understood to arise as a consequence of the destabilisation of the CE phase. It is known that the CE phase is made up by the antiferromagnetic allignment of one dimensional ferromagnetic zig zag chains. The orbital ordering (and consequently the insulating nature) arises due to the presence of staggered phase factors as the charge carriers hop along this chain.\cite{khomskii} Introduction of Al substitution disrupts the orbital ordering, thus making hopping more favourable. This is alo proved by the fact that Mn site substitution is known to result in an appreciable decrease in the resistivity of these CE systems,\cite{sunil,damay} and in the case of magnetic substution can even result in a metallic ground state due to dopant mediated ferromagnetic interactions. Since the extent of EPS depends crucially on the mobility of charge carriers, an increase in the hopping probablity would make it more easier for the carrier to equilibrate faster, thus making EPS correspondingly more difficult. However, it would be erronous to conclude that there is no EPS in the samples with higher disorder based on our observations (the T${_N}$ in these samples is till far lower than what would be expected with site dilution and finite hysteresis is observed right upto the 10\% doped sample); it only implies that more sensitive experiments on preferably shorter time scales would be required to unambigously detect the dynamics of EPS in these systems. 
 
In summary, based on our experimental observations, we provide a reasoning for the abnormally large sensitivity of the CE phase to quenched disorder. The rapid supression of T${_N}$ in this phase occurs as a consequence of finite size effects in these samples, which arises due to the stabilisation of electronically phase separated self organised structures. Quenched disorder stabilises cluster formation and the correlation length is limited by the size of these clusters. Though the quenched disorder triggers the stabilisation of cluster formation, EPS is the main driving force of this state. This low temperature metastable state is associated with large relaxation times and gives rise to striking magnetic memory effects as a function of the field and temperature measurement protocols. Increasing disorder reduces the extent of EPS, since the disruption of the CE phase increases the effective hopping probablity of the charge carriers, thus allowing them to equilibriate faster on account of their enhanced mobility.

\end{document}